\documentclass{book}    
\usepackage{piers}  
\usepackage{epstopdf}
\DeclareGraphicsExtensions{.pdf,.eps,.png,.jpg,.mps}
\begin{document}

\title{Simulation of motional averaging using a superconducting circuit}
\maketitle

\author      {J. Li}
\affiliation {O.V. Lounasmaa Lab}
\address     {}
\city        {Espoo}
\postalcode  {}
\country     {Finland}
\phone       {345566}    
\fax         {233445}    
\email       {email@email.com}  
\misc        { }  
\nomakeauthor

\author      {F. M. Lastname}
\affiliation {University}
\address     {}
\city        {Boston}
\postalcode  {}
\country     {USA}
\phone       {345566}    
\fax         {233445}    
\email       {email@email.com}  
\misc        { }  
\nomakeauthor

\begin{authors}
{\bf J. Li}$^{1}$, {\bf M. P. Silveri}$^{2}$, {\bf K. S. Kumar}$^{1}$, {\bf J.-M. Pirkkalainen}$^{1}$,  {\bf A. Veps\"al\"ainen}$^{1}$, {\bf W. C. Chien}$^{3}$, {\bf J. Tuorila}$^{2}$, {\bf M. A. Sillanp\"a\"a}$^{4}$,
{\bf P. J. Hakonen}$^{1}$, {\bf E. V. Thuneberg}$^{2}$, {\bf and G. S. Paraoanu}$^{1}$
 \\

\medskip
$^{1}$O.V. Lounasmaa Laboratory, Aalto University, PO Box 15100, FI-00076 AALTO, Finland\\
$^{2}$ Department of Physics, University of Oulu, P.O. Box 3000, FI-90014, Finland\\
$^{3}$ Department of Physics, National Chung Hsing University, 250 Kuo Kuang Road, Taichung 40227, Taiwan\\
$^{4}$ Department of Applied Physics, Aalto University, FI-00076 AALTO, Finland\\
\end{authors}

\begin{paper}

\begin{piersabstract}

The possibility of using a quantum system to simulate another one has been recognized for a long time as an important research direction in quantum information and quantum computing. In Ref. \cite{Li}, a superconducting circuit (a transmon) was employed to simulate a NMR (nuclear magnetic  resonance) effect known as motional averaging. In this paper we analyze the results of numerical integration of the time evolution of the density matrix of a qubit with random frequency fluctuations, and compare the results with those obtained by using the method of quantum trajectories. We show that both methods produce equivalent results, although some differences may appear in the range of intermediate modulation frequencies.

\end{piersabstract}

\psection{Introduction}

The generic problem of frequency-modulation of the characteristic or proper frequency of a physical system appears in a vast variety of contexts. In the field of superconducting circuits \cite{nori,Wendin}, frequency modulation is typically realized by changing the magnetic field that penetrates a SQUID (superconducting interference device) loop. One phenomenon associated with the change in frequency is parametric amplification, either of a classical input signal or of vacuum fluctuations \cite{Pasi}. Another important phenomenon is motional averaging and narrowing. This phenomenon has been first observed \cite{Bloembergen} and described theoretically \cite{Anderson} in ensembles of nuclei measured by NMR (nuclear magnetic resonance) techniques. Today it is one of the textbook results in the field of NMR \cite{Abragam}. More recently, the phenomenon of motional averaging has been demonstrated with the protons in water molecules \cite{Kohmoto}, and in ensembles of two-level systems consisting of ultracold $^{87}$Rb atoms \cite{Sagi}.

In contrast to the experiments involving ensembles of particles, our realization of motional averaging \cite{Li} uses a single quantum system, whose transition frequency between the ground state and the first excited state can be controlled externally. This system is a superconducting circuit consisting of a capacitively-shunted charge qubit (a transmon) embedded in a superconducting waveguide resonator \cite{koch}. Two microwave fields are applied to the system: one, at the frequency of the resonator, is used for measurement, while the other, around the qubit frequency, is used for driving. In addition, the qubit frequency is modulated by a random telegraph noise with externally-controlled amplitude and characteristic jumping frequency $\chi$, defined more precisely below. Surprisingly, by adding noise in this way a new, ``motional averaged'' spectral line is formed, with a linewidth smaller than the amplitude of the random modulation \cite{Li}. We have also succeeded in driving Rabi oscillations on the motional averaged line, demonstrating the formation of hybrid states of the transmon and the modulation field. The fact that quantum coherence persists in the presence of noise and that a spectral line appears at an average frequency where  - at least ideally -  the qubit only crosses very fast without spending any time into is a rather counterintuitive feature of this effect, adding up to the list of other unexpected quantum effects that were demonstrated in recent times with superconducting qubits, such as interaction-free \cite{free} and partial \cite{partial} measurements, reversal of nonunitary transformations \cite{reversal}, and the violation of Leggett-Garg inequalities \cite{violation}.

When modulating the system sinusoidally, we observe a rich spectral structure, resembling a Landau-Zener interference pattern. However, direct Landau-Zener transitions are prohibited in our system by the fact that the frequency of the modulation is much smaller than the energy level separation. We were able to show that in this case the transitions occur through the absorption of photons from the driving field; we call this process {\bf photon-assisted Landau-Zener effect}.

Finally, for the values of the fields used in our experiment we show that, in a rotating frame, the system reaches the ultrastrong coupling regime. Our setup can also be seen as the simulation of the effect of coupling an externally-controlled fluctuation to a qubit. We have also suggested that the experimental demonstration of motional averaging shown here could provide a novel route to improving the dephasing times of existing superconducting qubits.

In this contribution we analyze the simplest theoretical model for motional averaging, consisting of a frequency-modulated two-level system under driving and with dissipation. We present the results of two methods used for calculating the time-evolution of this system, namely direct numerical integration and the method of quantum trajectories. The two methods produce overall the same spectra, but for certain parameter choice some differences may appear in the regime of intermediate modulation frequencies.

\psection{Model and results}

We consider a generic qubit Hamiltonian with bare (unmodulated) frequency $\omega_{0}$,
\begin{equation}
H_{0}=\frac{\hbar}{2}\omega_{0}\sigma_{z},
\end{equation}
on top of which we add a frequency-modulated term $\xi(t)$, resulting in a total time-dependent Hamiltonian
\begin{equation}
H (t)=\frac{\hbar}{2}\left[\omega_{0}+\xi (t)\right]\sigma_{z}.
\end{equation}
For the modulation, we consider a stochastic Poisson process in which $\xi (t)$ switches randomly between two values, $\pm\xi$, with mean jumping frequency denoted by $\chi$. For such processes, the probability of $n$ jumps within a time interval $t$ is given by
\begin{equation}
P_{n}(t) = \frac{1}{n!}(\chi t)^{n}e^{-\chi t}.
\end{equation}
The correlations between the jumping events at times $t_n$ are maximally anti-correlated, namely $\langle \xi (t_{n+1})\xi ({t_{n}})\rangle = -\xi^{2}$.
In addition, the qubit is driven by a $\sigma_{x}$-coupled tone with angular frequency $\omega$,
\begin{equation}
H_{\rm drive}(t) = \hbar g  \cos(\omega t)\sigma_{x} ,
\end{equation}
where $g$ is the strength of the coupling (the Rabi frequency). It is useful to truncate the driving Hamiltonian to energy-conserving terms, by employing a rotating-wave approximation,
\begin{equation}
H_{\rm drive}(t) \approx H_{\rm drive}^{\rm RWA} (t)= \frac{\hbar g}{2}  \left[\sigma e^{i\omega t}+ \sigma^{+}e^{-i\omega t}\right].
\end{equation}
Thus the total Hamiltonian in the rotating-wave approximation reads
\begin{equation}
H_{\rm tot}^{\rm RWA} (t)= H_{\rm drive}^{\rm RWA} (t) + H (t).
\end{equation}
Finally, dissipation is introduced via the standard Liouvillean approach to superconducting qubits, see {\it e.g.} \cite{Jian_prb}. The time-dependent equation satisfied by the density matrix $\rho$ takes the form
\begin{equation}
\dot{\rho}(t) = -\frac{i}{\hbar}\left[H_{\rm tot}^{\rm RWA}(t), \rho (t)\right] + {\cal L}[\rho (t)],\label{eq}
\end{equation}
where the Liouvillean superoperator at zero temperature is
\begin{equation}
{\cal L}[\rho (t)] = \frac{\Gamma_1}{2} [2\sigma \rho (t) \sigma^{+} - \sigma^{+}\sigma\rho (t) - \rho (t)\sigma^{+}\sigma ]
+\frac{\Gamma_{\varphi}}{2}[\sigma_{z}\rho (t)\sigma_{z} - \rho (t)].
\end{equation}
Here $\sigma$ and $\sigma^{+}$ are the lowering and raising Pauli matrices, $\Gamma_1$ is the relaxation rate, and $\Gamma_\varphi$ is the dephasing rate. The Liouvillean
can be written in the matrix form as
\begin{equation}
{\cal L}[\rho (t)] = \frac{1}{2}\left[\begin{matrix} 2\Gamma_{1}\rho_{11}(t) & -(\Gamma_{1} + 2\Gamma_{\varphi})\rho_{01}(t) \\
-(\Gamma_{1} + 2\Gamma_{\varphi})\rho_{10}(t) & -2\Gamma_{1}\rho_{11}(t)\end{matrix}\right].
\end{equation}
Eq. (\ref{eq}) can be directly solved numerically, by using the standard Runge-Kutta method \cite{Jian_thesis}. In contrast, in \cite{Li} we have solved the evolution by using the quantum trajectories technique \cite{trajectories}, see \cite{Matti_thesis}. We have run in parallel several simulations using these methods, demonstrating that they yield very similar-looking spectra. There are however some interesting differences, that appear more visibly for a certain range of parameters. Here we present such an example.

In Fig. 1 we show the result of simulations using the two methods described above. The direct integration technique using the standard Runge-Kutta method for solving differential equations produces the spectrum shown in the right plot of Fig. 1. Interestingly, this shows that the two lines in the spectrum at low jumping frequencies spread up a bit when $\chi$ increases, before collapsing into the motional-averaged line. Also, the probability of qubit excitation at frequencies between the two initial values $\pm\xi = \pm 71$ MHz tends to be larger. This can be understood as coming from multiple-photon processes which are not completely extinct as one would expect for ideal random pulses. Indeed, one sees clearly these multiple-photon spectral lines when modulating  the qubit frequency with a sine wave \cite{Jian_thesis}. An ideal random telegraph-noise pulse can be expanded in an infinite number of  frequency components, each of them producing a set of sidebands. Since they are at different frequencies, all these sidebands will average out, with the exception of the middle one (the zero-photons sideband), which is common to all frequencies and which produces the motional-averaged line. Interestingly, the feature described above (slight up-spreading of the two spectral lines) appears also in some of the experimental data, see \cite{Jian_thesis}. The advantage for using the method of quantum trajectories is that the computer runtime is much shorter, since it can be easily parallelized.
The simulation for a 50 ($x$-axis)$\times$ 101 ($y$-axis) points and using the
mirror symmetry with respect to the $\omega=\omega_0$ -line, as presented here, would have taken 378 hours
for a single CPU but by parallelizing for 21 CPUs it took only approx 18
hours. The number of averaged traces was 1000.

\begin{figure}[!h]
\begin{center}
\includegraphics[width=8cm]{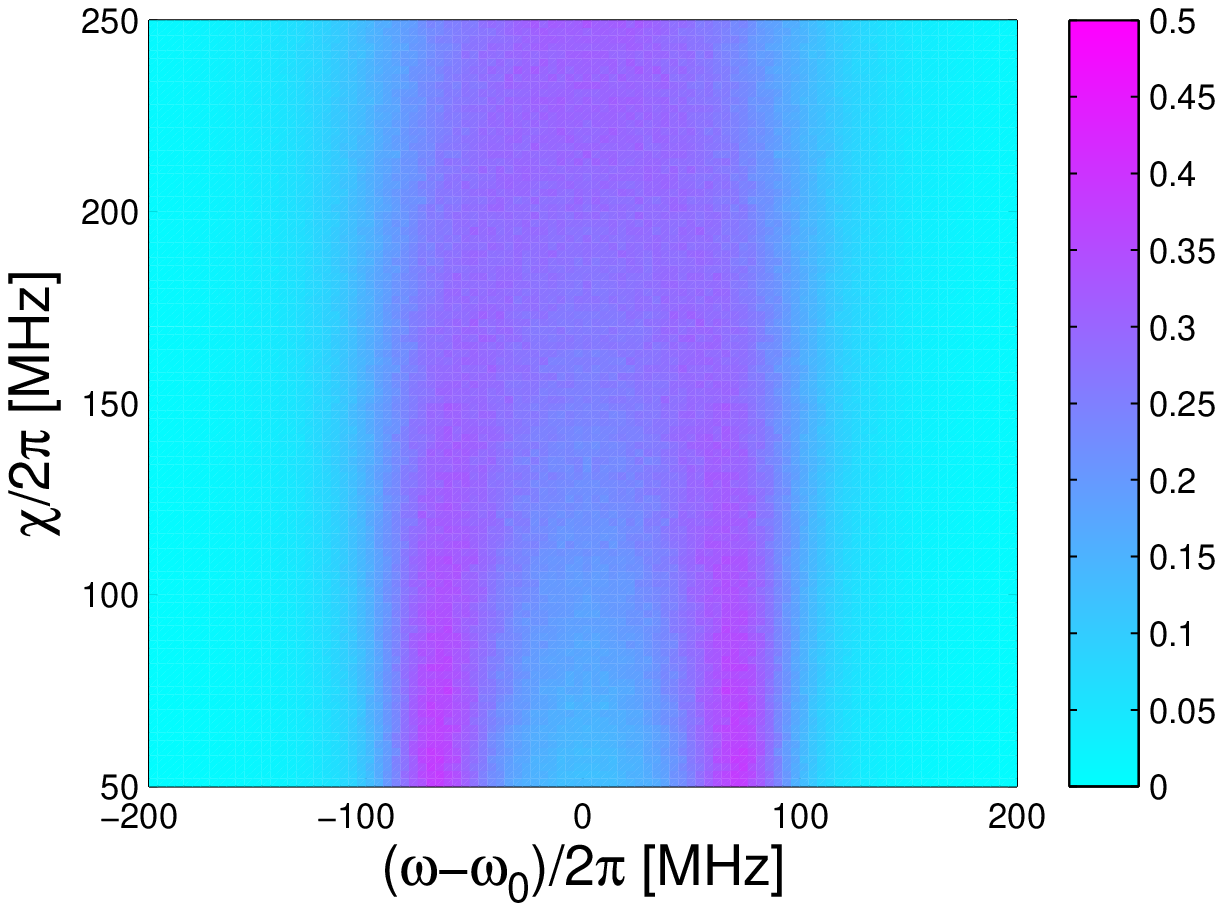}
\includegraphics[width=8cm]{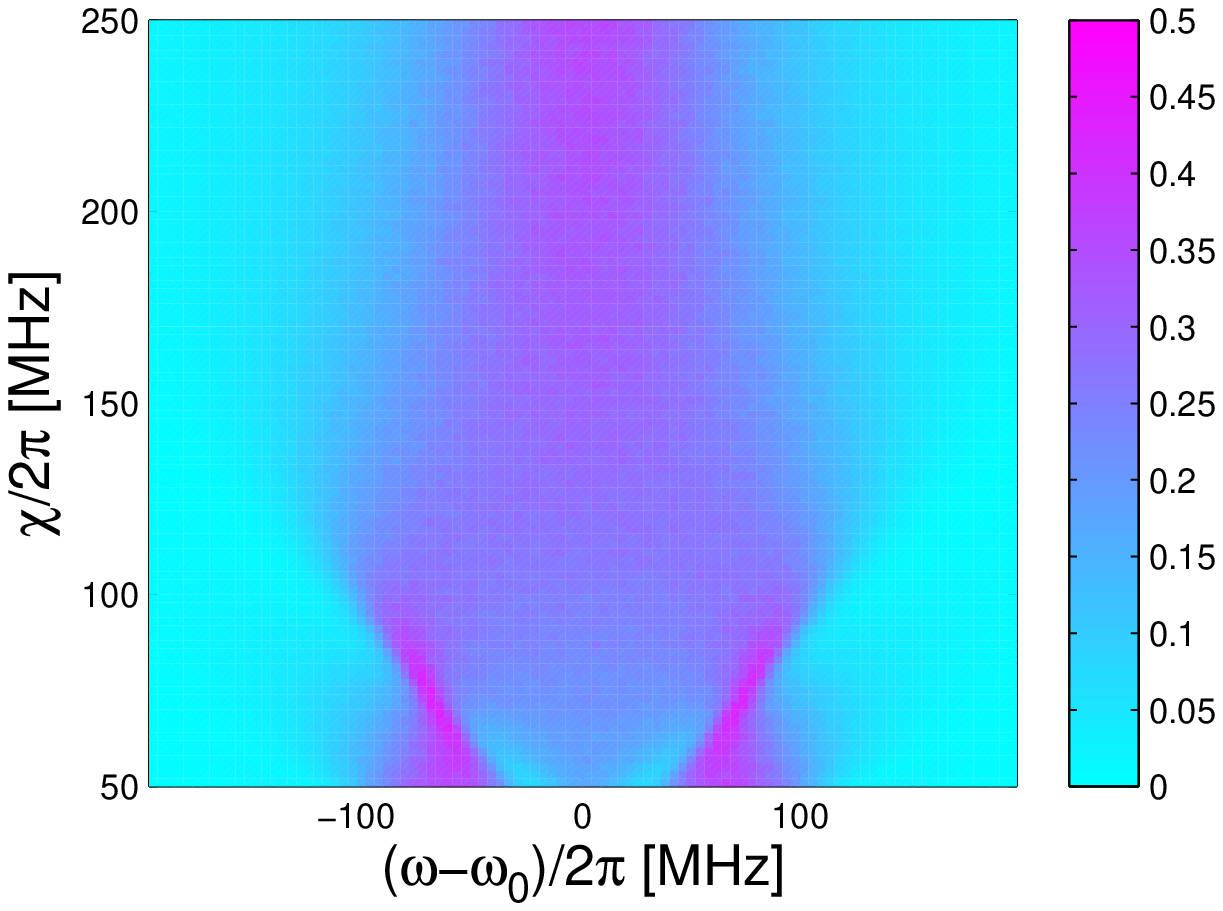}
\end{center}\label{onefig}
\caption{Results of simulations for the qubit population for $\pm\xi=71$ MHz, $\Gamma_{1} = 2\pi \times 1$ MHz, $\Gamma_\varphi = 2\pi \times 2.5$ MHz
Left plot: simulation using  the method of quantum trajectories. The raise/fall time was 2 ns, and it was sampled in three intermediate steps, each of them with width of 0.5 ns. The figure is obtained with 1000 averages. Right plot: simulation by direct numerical integration using the Runge-Kutta method, with 100 averages.  The rise/fall time was 1.6 ns, and the width of the time step was 0.1 ns. The two spectra are quite similar, but there are some differences in the intermediate jump frequency regime. The direct simulation partly retains the multiple-photon sidebands, which, at low jumping frequencies, tend to blur the space between the two jump frequencies, and also somewhat extend the two spectral lines sideways. These features are almost completely washed out in the left plot. At high jump frequencies, the results of these simulations are nearly identical.}
\end{figure}

\psection{Summary and conclusion}

When the frequency of an artificial atom realized as a superconducting qubit is varied fast enough by the application of a sequence of random pulses, a single line emerges in the absorbtion spectrum. This phenomenon is known as motional averaging.
Since in general the linewidth of quantum systems is determined by such random processes ({\it e.g.} 1/f noise), this process is often referred to as motional narrowing, since it appears as the narrowing of the spectral line when these processes are fast enough. Our experiment can be seen as simulating the above phenomenon using a superconducting cicuit. In this contribution we have discussed the numerical predictions for the spectra using two different techniques, one which is a direct numerical calculation and the other which employs the method of quantum trajectories. We found that the two methods give equivalent results, showing the consistency of these theoretical approaches. The differences between the two can be explained by the incomplete averaging out of multiple-photon sidebands in the case of the direct integration method. Also, the second method is faster and can be easily parallelized.

\ack
The work was supported by the Academy of Finland, through projects 135135, 141559, 263457 and the Center of Excellence ``Low Temperature Quantum Phenomena and Devices'' project 250280. W. C. Chien was supported from Academy of Finland project 253094, related to the cooperation between Finland and Taiwan. G. S. P. would like to thank Watson Kuo for useful comments.

\end{paper}

\end{document}